\documentclass[conference]{IEEEtran}
\IEEEoverridecommandlockouts
\usepackage{cite}
\usepackage{amsmath,amssymb,amsfonts}
\usepackage{graphicx}
\usepackage{textcomp}
\usepackage{xcolor}
\usepackage{algorithm}   % Required for the algorithm environment
\usepackage[utf8]{inputenc}

\usepackage{algorithmicx} % Core package for writing algorithms
\usepackage{algpseudocode} % Provides structured pseudocode
\usepackage{amsmath}
\usepackage{hyperref}
\def\BibTeX{{\rm B\kern-.05em{\sc i\kern-.025em b}\kern-.08em
    T\kern-.1667em\lower.7ex\hbox{E}\kern-.125emX}}

    \algrenewcommand\alglinenumber[1]{\scriptsize #1}
\begin{document}

\title{Smart Water Security with AI and Blockchain-Enhanced Digital Twins}

\author{\IEEEauthorblockN{1\textsuperscript{st} Mohammadhossein Homaei}
\IEEEauthorblockA{\textit{Media Engineering Group} \\
\textit{University of Extremadura}\\
C\'aceres, Spainn \\
mhomaein@alumnos.unex.es}
\and
\IEEEauthorblockN{2\textsuperscript{nd} Víctor González Morales }
\IEEEauthorblockN{3\textsuperscript{rd} Óscar Mogollón Gutiérrez}
\IEEEauthorblockA{\textit{Media Engineering Group} \\
\textit{University of Extremadura}\\
C\'aceres, Spain \\
\{victorgomo, oscarmg\}@unex.es}
\and
\IEEEauthorblockN{4\textsuperscript{th} Rubén Molano Gómez}
\IEEEauthorblockN{5\textsuperscript{th} Andrés Caro}
\IEEEauthorblockA{\textit{Media Engineering Group} \\
\textit{University of Extremadura}\\
C\'aceres, Spain \\
\{rmolano, andresc\}@unex.es}
}

\maketitle

\begin{abstract}
Water distribution systems in rural areas face serious challenges such as a lack of real-time monitoring, vulnerability to cyberattacks, and unreliable data handling. This paper presents an integrated framework that combines LoRaWAN-based data acquisition, a machine learning-driven Intrusion Detection System (IDS), and a blockchain-enabled Digital Twin (BC-DT) platform for secure and transparent water management. The IDS filters anomalous or spoofed data using a Long Short-Term Memory (LSTM) Autoencoder and Isolation Forest before validated data is logged via smart contracts on a private Ethereum blockchain using Proof of Authority (PoA) consensus. The verified data feeds into a real-time DT model supporting leak detection, consumption forecasting, and predictive maintenance. Experimental results demonstrate that the system achieves over 80 transactions per second (TPS) with under 2 seconds of latency while remaining cost-effective and scalable for up to 1,000 smart meters. This work demonstrates a practical and secure architecture for decentralized water infrastructure in under-connected rural environments.
\end{abstract}


\section{Introduction}\label{introduction}

While remote-sensing techniques have proven valuable for water quality monitoring \cite{cuartero2023, caceres2024}, regarding water distribution, efficient distribution is a significant issue in rural regions, especially where infrastructure is poor and digital monitoring is scarce. Many rural parts of Spain still rely on outdated water distribution systems with manual inspections or partially automated tools, leading to delays in problem detection, inaccurate usage data, and increased risks of human error or data manipulation. Implementing a digital twin (DT) can address these challenges by creating a virtual replica of the water network, enabling operators to detect leaks, predict demand, and optimize maintenance scheduling. However, DT systems also face cybersecurity risks, as data may be altered or falsified before reaching the digital model \cite{Alshami2024, Homaei2024}.

DT technology is increasingly central to cyber-physical systems. It provides a real-time virtual representation of physical infrastructures, supporting advanced analysis, forecasting, and anomaly detection \cite{Li2024, Krishnan2022, Homaei2022}. DTs assist utility operators in identifying leaks, pressure irregularities, and unusual consumption patterns. Despite these advantages, ensuring data security and trustworthiness within DT-based systems remains a significant challenge, particularly in decentralized settings involving multiple stakeholders.

BC technology offers a novel solution for secure data management within online platforms. It ensures decentralized and tamper-proof information storage and verification through Distributed Ledger Technology (DLT), cryptographic security, and consensus mechanisms \cite{Fuller2020, Homaei2024}. Additionally, smart contracts automate tasks like device registration, real-time billing, and fault detection, thus minimizing reliance on intermediaries and increasing accountability \cite{Kirli2022, Satilmisoglu2024}.

To address these security concerns, our platform incorporates an Artificial Intelligence (AI)-based IDS that identifies anomalous or suspicious data. Only validated and trusted data are forwarded to the BC, where they are securely and permanently stored. Our method integrates LSTM-based anomaly detection and BC technology, ensuring precise, secure, and transparent water management. Without secure, intelligent prediction, water use remains inefficient and decisions are delayed.

This paper presents an integrated DT system supported by BC technology to enhance water distribution management in rural villages in Spain. Data is collected via long-range, low-power LoRaWAN sensors and securely stored on a private BC network, creating an encrypted foundation for the DT. AI and ML techniques enable predictive analytics and anomaly detection, supporting better decision-making and resource optimization. The integration of BC and DT technologies enhances transparency, scalability, and reliability. This cost-effective solution is suitable for rural communities, water utilities, and governmental entities.

The remainder of this paper is structured as follows: Section~\ref{secII} reviews related studies on DT and BC applications in water management. Section~\ref{secIII} describes the proposed framework, including DT architecture, LoRa-based sensor networks, and the integration of a private BC. Section~\ref{secIV} evaluates performance, scalability, and security through real-world tests. Section~\ref{secV} discusses key findings, challenges, and implications, and concludes by highlighting the contributions and suggesting future research directions aimed at optimizing BC-based DT solutions.

\section{Related Works}\label{secII}

\subsection{DT in the Water Industry}
Recent studies have highlighted the role of DTs as effective tools for enhancing water distribution systems. DT technologies simulate real-time water network behavior, enabling operators to perform leak detection, forecast water consumption, and improve overall maintenance scheduling \cite{Homaei2025, Li2024}. For instance, DT platforms have successfully reduced water losses through predictive analytics, proving valuable for improving efficiency and reducing operational costs \cite{Krishnan2022}. However, existing DT implementations often neglect critical cybersecurity considerations, treating data from IoT sensors as inherently trustworthy, which can expose systems to data spoofing and manipulation risks \cite{Alshami2024}.

\subsection{Security in Water Distribution Systems}

The digitalization of water systems introduces increased cybersecurity threats, particularly where IoT devices and wireless sensor networks (e.g., LoRaWAN) are extensively deployed. Kim et al. \cite{Kim2021water} revealed multiple vulnerabilities, including weak encryption methods, poor authentication practices, and outdated communication protocols, leaving these systems susceptible to unauthorized access, data falsification, and denial-of-service (DoS) attacks. Traditional cybersecurity frameworks for water infrastructure tend to reactively detect breaches only after data has been compromised or corrupted, lacking real-time predictive detection capabilities \cite{Park2023}. Consequently, an urgent need exists for proactive anomaly detection systems integrated seamlessly within digital water infrastructures to protect against threats before data reaches critical operational layers like DTs.

\subsection{BC Integration in DT}

BC technology has been proposed as a viable solution to improve security, transparency, and immutability in DT applications across multiple sectors. For instance, Mohammed et al. \cite{Mohammed2024} employed Hyperledger Fabric to secure IoT sensor data in smart water management systems, demonstrating enhanced trust and traceability. Similarly, MQTT-based BC solutions have successfully provided tamper-proof logging of sensor readings, reducing risks of data loss and falsification \cite{Naqash2023}. Despite these advances, existing BC-integrated DT frameworks assume incoming data is valid without verifying it, creating vulnerabilities. Teisserenc et al. \cite{Teisserenc2021} addressed some limitations by introducing decentralized DT models with smart contracts for automated decision-making, but lacked robust pre-validation mechanisms to ensure data authenticity. Thus, incorporating anomaly detection before BC data storage is essential to ensure data integrity.

\subsection{AI and ML in IDS for Critical Infrastructure}
To overcome limitations of traditional security approaches, AI and ML techniques have become widely adopted for anomaly detection in critical infrastructure, such as energy grids and water systems. Isolation Forest algorithms have successfully identified statistical anomalies in infrastructure sensor data, enabling early detection of attacks and faults \cite{Fuller2020}. Additionally, LSTM Autoencoders have proven effective in detecting temporal anomalies in sequential data, such as abnormal water consumption patterns indicative of leaks or cyberattacks \cite{Kirli2022, Gutierrez2022}. Nevertheless, existing ML-based IDS solutions are often developed in isolation, lacking integrated deployment within BC-enabled DT frameworks. Furthermore, their evaluation typically occurs under idealized laboratory conditions, without sufficient consideration of intermittent connectivity or rural operational constraints. This research gap motivates the integration of AI-driven IDS within BC and DT ecosystems, specifically addressing rural deployments.

\begin{figure}
\centering
\includegraphics[width=1\linewidth]{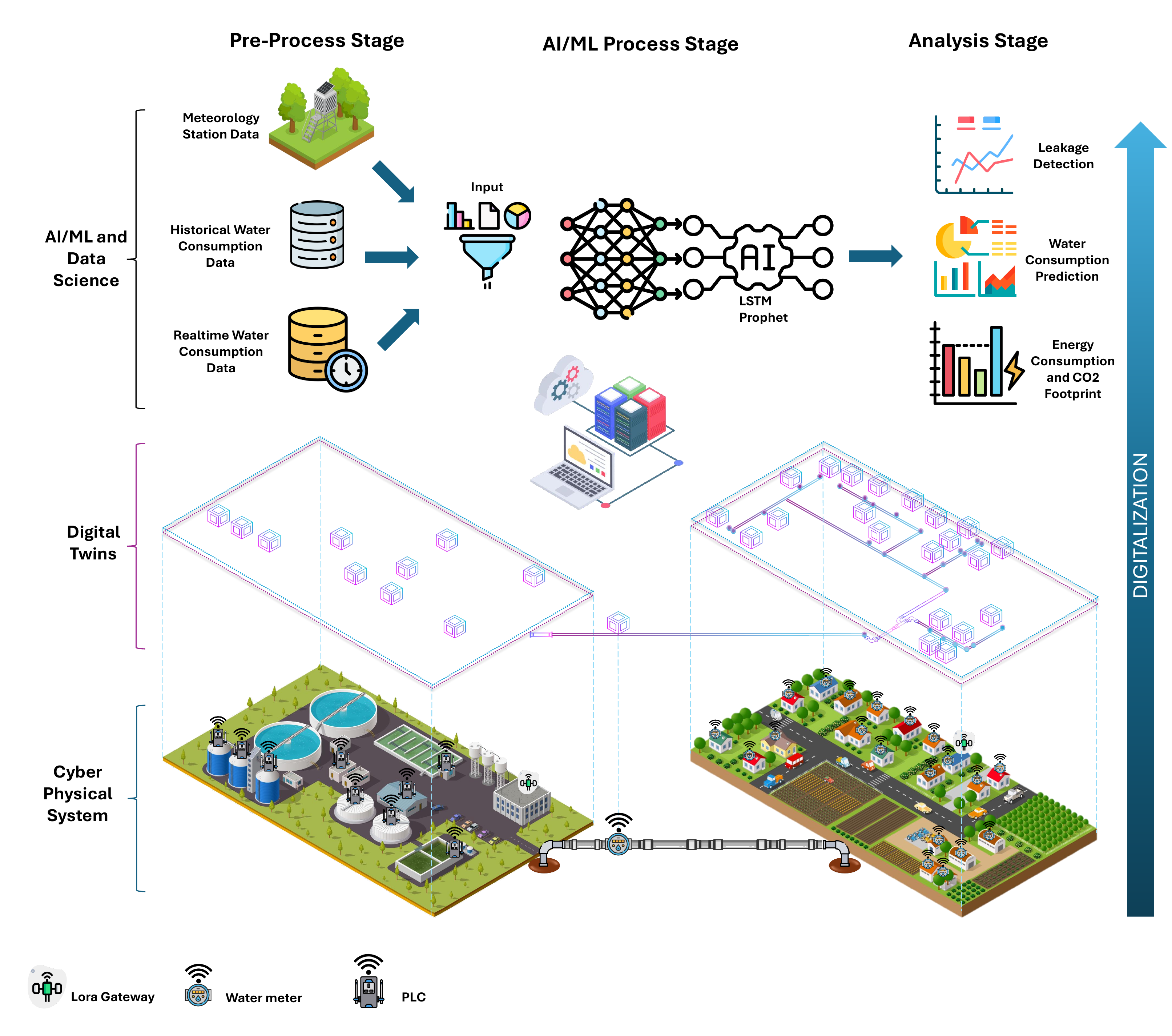}
\caption{A Digital Twin Platform in the Water Industry~\cite{Homaei2025}}
\label{fig:DT_platform}
\end{figure}

\section{Proposed Platform }\label{secIII}
In this paper, we build on our previous work in the water industry by enhancing the DT system with a stronger focus on security, reliability, and data protection. The DT model proposed in \cite{Homaei2025} is composed of three main layers (Figure~\ref{fig:DT_platform}): the cyber-physical system (CPS), its digital representation, and an AI-based data analysis and prediction layer. In this updated version, we improve the leak detection process, simplify the identification of data anomalies, and leverage historical data to detect potential cyberattacks and abnormal patterns. To enhance transparency, we integrate a BC system that connects LoRaWAN sensor data to a private Ethereum-based BC. The improved system consists of three key components:

\begin{itemize}
    \item Leak and unreal detection layer, which processes incoming data and compares it against seasonal trends and long-term average consumption patterns.
    \item Anomaly and attack detection layer, based on an LSTM model, which prevents out-of-range or incorrect data from being inserted into the BC.
    \item BC layer, which securely stores validated data from the previous layers on the BC using a dedicated smart contract.
\end{itemize}

\subsection{Leakage detection}
The first step in ensuring water system integrity involves detecting leaks and unrealistic consumption values based on temporal patterns. The platform uses rule-based logic and thresholding derived from long-term consumption profiles to identify potential leaks, especially during non-usage hours (e.g., 00:00–06:00). The logic assumes that consistent water flow during expected inactivity periods likely indicates leakage. Algorithm~\ref{alg:leakage-detection} summarizes this detection process, which serves as a lightweight filtering mechanism before invoking the AI-based IDS for further validation. Leakage alerts are issued locally and passed to the IDS for anomaly confirmation before BC logging.

\begin{algorithm}
\caption{Leakage Detection and Blockchain Validation}
\label{alg:leakage-detection}
\begin{algorithmic}[1]
\scriptsize
\State \textbf{Input:} LoRaWAN meter data stream $D$
\State \textbf{Output:} Leakage alerts, validated BC records
\State Initialize buffer $H$, counters
\ForAll{$d \in D$}
    \State Extract hourly data and update $H$
    \State Check nighttime (00:00--06:00) consumption
    \State Update leakage counter: increment if all $>0$, else reset
    \If{counter $\geq 2$}
        \State Flag leakage; freeze status
        \If{next message confirms leakage} \State Alert consumer \EndIf
    \EndIf
    \State Run IDS on $d$
    \If{anomaly detected} \State Log and reject \Else \State Store on BC \EndIf
\EndFor
\end{algorithmic}
\end{algorithm}

\subsection{AI-Based IDS for DT}
While the BC component guarantees secure and immutable data storage, it does not provide real-time protection against data spoofing, replay attacks, or transaction flooding. To address this, we propose an AI-driven IDS that operates between the data acquisition and BC layers. This IDS employs two complementary models: an LSTM Autoencoder for sequence-based anomaly detection and an Isolation Forest (IF) for statistical outlier detection. Together, they filter out both temporal and point-wise anomalies in water meter data received via LoRaWAN before storing it on the BC.

\subsubsection{Design and Threat Model}
The IDS is designed to detect and block:
\begin{itemize}
    \item \textit{Spoofed data:} Manipulated readings with plausible structure but inconsistent behavior.
    \item \textit{Replay attacks:} Repetition of legitimate data to flood or mislead the system.
    \item \textit{Outliers:} Abnormally high consumption, unexpected error codes, or gas usage irregularities.
\end{itemize}
While smart contracts verify sender identity and enforce structural rules, they cannot detect logical inconsistencies. The IDS addresses this gap through both statistical and temporal pattern learning.

\subsubsection{Feature Engineering}
The IDS continuously processes incoming real-time data streams from LoRaWAN sensors and BC logs. For each new record, a feature vector is constructed as:
\begin{equation}
\mathbf{x}_t = \left[\text{WaterUsage}_t, \text{ErrorCode}_t, \text{TxRate}_t, \text{GasUsed}_t\right]
\end{equation}
For the LSTM model, a sequence of $N$ such vectors is maintained per meter:
\begin{equation}
\mathbf{X}_m = \{\mathbf{x}_{t-N+1}, \ldots, \mathbf{x}_t\}
\end{equation}

\begin{algorithm}[!t]
\caption{Combined LSTM and Isolation Forest IDS}
\label{alg:lstm_if_ids}
\begin{algorithmic}[1]
\scriptsize
\State \textbf{Input:} Trained LSTM model $\mathcal{M}$, trained IF model $\mathcal{F}$, thresholds $\tau$, $\theta$
\State \textbf{For each meter:} maintain buffer $\mathbf{X}_m$ of size $N$
\ForAll{incoming event $e_t$}
    \State Extract feature vector $\mathbf{x}_t$ and append to $\mathbf{X}_m$
    \State Compute IF anomaly score: $s \gets \mathcal{F}(\mathbf{x}_t)$
    \If{$s > \theta$}
        \State Raise anomaly alert (Isolation Forest)
        \State Reject record and log the incident
    \ElsIf{$|\mathbf{X}_m| = N$}
        \State $\hat{\mathbf{X}}_m \gets \mathcal{M}.decode(\mathcal{M}.encode(\mathbf{X}_m))$
        \State Compute reconstruction loss $\mathcal{L}_{\text{recon}}$
        \If{$\mathcal{L}_{\text{recon}} > \tau$}
            \State Raise anomaly alert (LSTM Autoencoder)
            \State Reject record and log incident
        \Else
            \State Accept and forward to BC
        \EndIf
        \State Remove oldest vector from $\mathbf{X}_m$
    \EndIf
\EndFor
\end{algorithmic}
\end{algorithm}

\subsubsection{LSTM Autoencoder Architecture}
The LSTM autoencoder learns to reconstruct sequences of normal behavior. It encodes a sequence into a latent representation and reconstructs it, allowing anomaly detection via reconstruction error:
\begin{equation}
\mathcal{L}_{\text{recon}} = \frac{1}{N} \sum_{i=1}^{N} \left\| \mathbf{x}_i - \hat{\mathbf{x}}_i \right\|^2
\end{equation}
If $\mathcal{L}_{\text{recon}} > \tau$, where $\tau$ is a predefined threshold, the sequence is flagged as anomalous.

\subsubsection{Isolation Forest Outlier Detection}
To complement the LSTM, we use an Isolation Forest trained on individual features to detect non-sequential outliers~\cite{Downey2024}. Given a new observation $\mathbf{x}_t$, the IF model returns an anomaly score $s(\mathbf{x}_t)$ based on how easily the point is isolated in the tree ensemble. An alert is raised if:
\begin{equation}
s(\mathbf{x}_t) > \theta
\end{equation}
where $\theta$ is the anomaly threshold determined during training.

\subsubsection{Real-Time Detection Algorithm}
The detection process operates in real time using a sliding window buffer and event-based triggers from smart contracts. A record must pass both LSTM-based sequence analysis and IF outlier detection before being accepted.

\subsubsection{Integration with Blockchain}
The IDS listens to smart contract events (e.g., \textit{WaterDataLogged}) via Web3 interfaces and buffers incoming records accordingly. Its output dictates whether the data is stored on the BC or discarded. Optional logging of detected anomalies on-chain can improve transparency, support audits, and train future models. The IDS layer is modular and can be deployed alongside any BC network or smart contract design, ensuring seamless integration and compatibility with decentralized infrastructures.

\subsection{BC}
\begin{figure}
    \centering
    \includegraphics[width=1\linewidth]{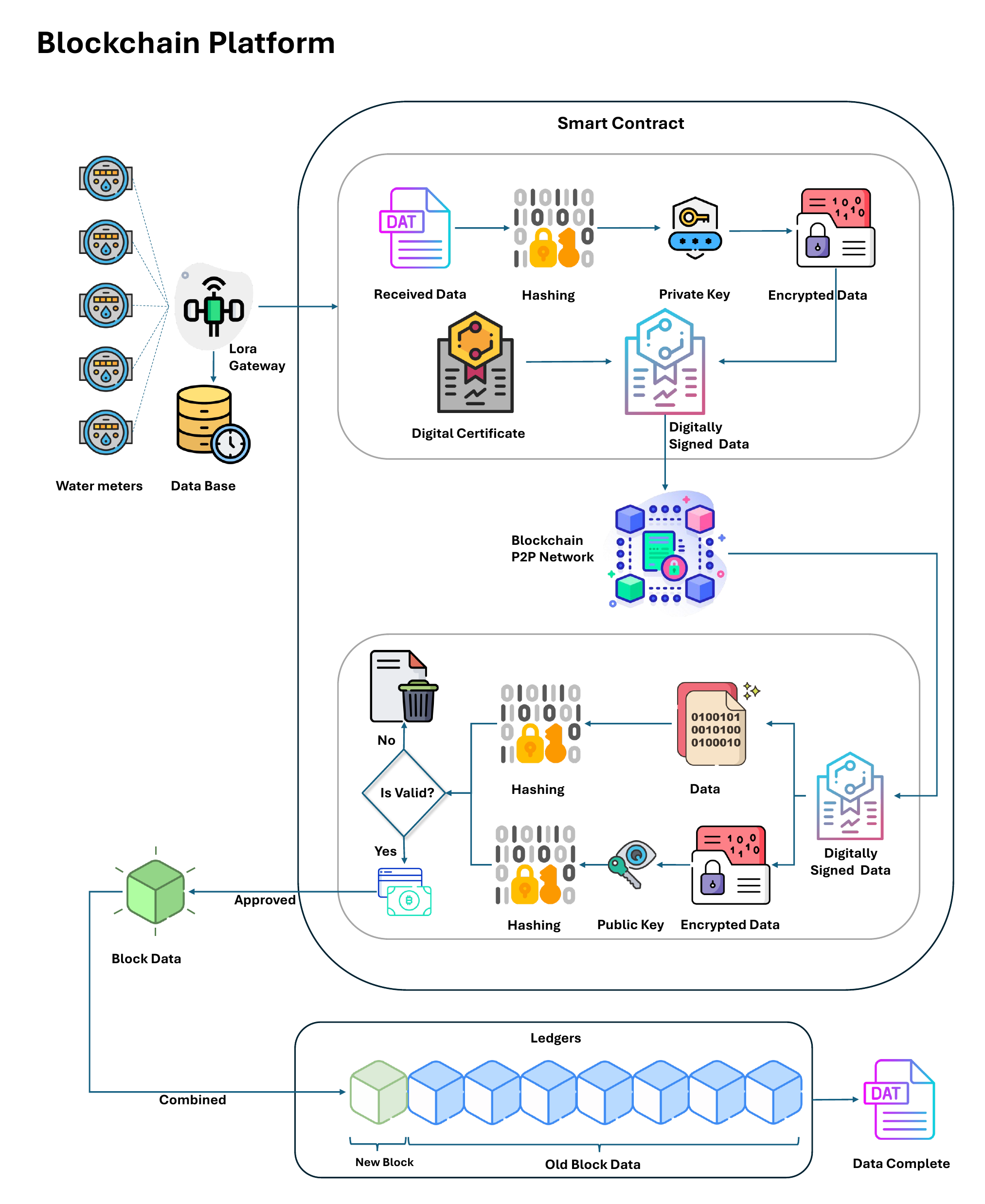}
    \caption{Proposed Smart contract for DT platform}
    \label{fig:blockchain-contract}
\end{figure}

Figure~\ref{fig:blockchain-contract} illustrates the smart contract structure used in the proposed platform. This contract handles essential functions such as meter registration, secure logging of consumption data, and automatic transaction processing. Specifically, the contract defines a `WaterData` structure that records the timestamp, water usage, error codes, and associated meter IDs. Functions such as \textit{registerMeter()}, \textit{logWaterData()}, and \textit{calculatePayment()} are designed to enforce access control, validate data, and automate billing based on consumption and error codes, as detailed in Algorithm~\ref{alg:smartcontract}.

\begin{algorithm}[!t]
\caption{DT and BC Smart Contract}
\label{alg:smartcontract}
\begin{algorithmic}[1]
\tiny

\State \textbf{Contract} VillageWaterSystem
\State \textbf{Struct} WaterData: \textbf{uint256} timestamp, waterUsage, errorCode; \textbf{string} meterId
\State \textbf{address} owner; \textbf{mapping(string => WaterData[])} waterLogs; \textbf{string[]} registeredMeters
\State \textbf{Event} MeterRegistered, MeterDisabled, WaterDataLogged, PaymentProcessed

\Function{onlyOwner}{}
    \State \textbf{Require}(msg.sender = owner, "Unauthorized")
\EndFunction

\Function{Constructor VillageWaterSystem}{}
    \State owner $\gets$ msg.sender
\EndFunction

\Function{registerMeter}{string meterId}
    \State \textbf{Require}(meterId $\neq$ empty, "Invalid ID")
    \State registeredMeters.push(meterId); \textbf{Emit} MeterRegistered(meterId)
\EndFunction

\Function{disableMeter}{string meterId}
    \State \textbf{Require}(meterId $\neq$ empty, "Invalid ID")
    \State Remove from registeredMeters; \textbf{Emit} MeterDisabled(meterId)
\EndFunction

\Function{logWaterData}{string id, uint256 u, uint256 e}
    \State \textbf{Require}(isMeterRegistered(id), "Unreg.")
    \State \textbf{Require}(e $\leq$ 100, "Invalid err")
    \State Store in waterLogs[id]; \textbf{Emit} WaterDataLogged(id, u, e)
    \State p $\gets$ calculatePayment(u, e); \textbf{Emit} PaymentProcessed(id, p)
\EndFunction

\Function{isMeterRegistered}{string id} \textbf{returns} bool
    \State \textbf{Return} (id in registeredMeters)
\EndFunction

\Function{calculatePayment}{uint256 u, e} \textbf{returns} uint256
    \State \textbf{Return} u * 1 ether * ((e $>$ 80) ? 1 : 2)
\EndFunction

\Function{getWaterLogs}{string id} \textbf{returns} WaterData[]
    \State \textbf{Require}(isMeterRegistered(id), "Unreg.")
    \State \textbf{Return} waterLogs[id]
\EndFunction

\Function{getRegisteredMeters}{} \textbf{returns} string[]
    \State \textbf{Return} registeredMeters
\EndFunction

\Function{setBaseRate}{uint256 r} onlyOwner
    \State \textbf{Require}(r $>$ 0, "Invalid rate")
\EndFunction

\end{algorithmic}
\end{algorithm}

\begin{figure}[!t]
    \centering
    \includegraphics[width=1\linewidth]{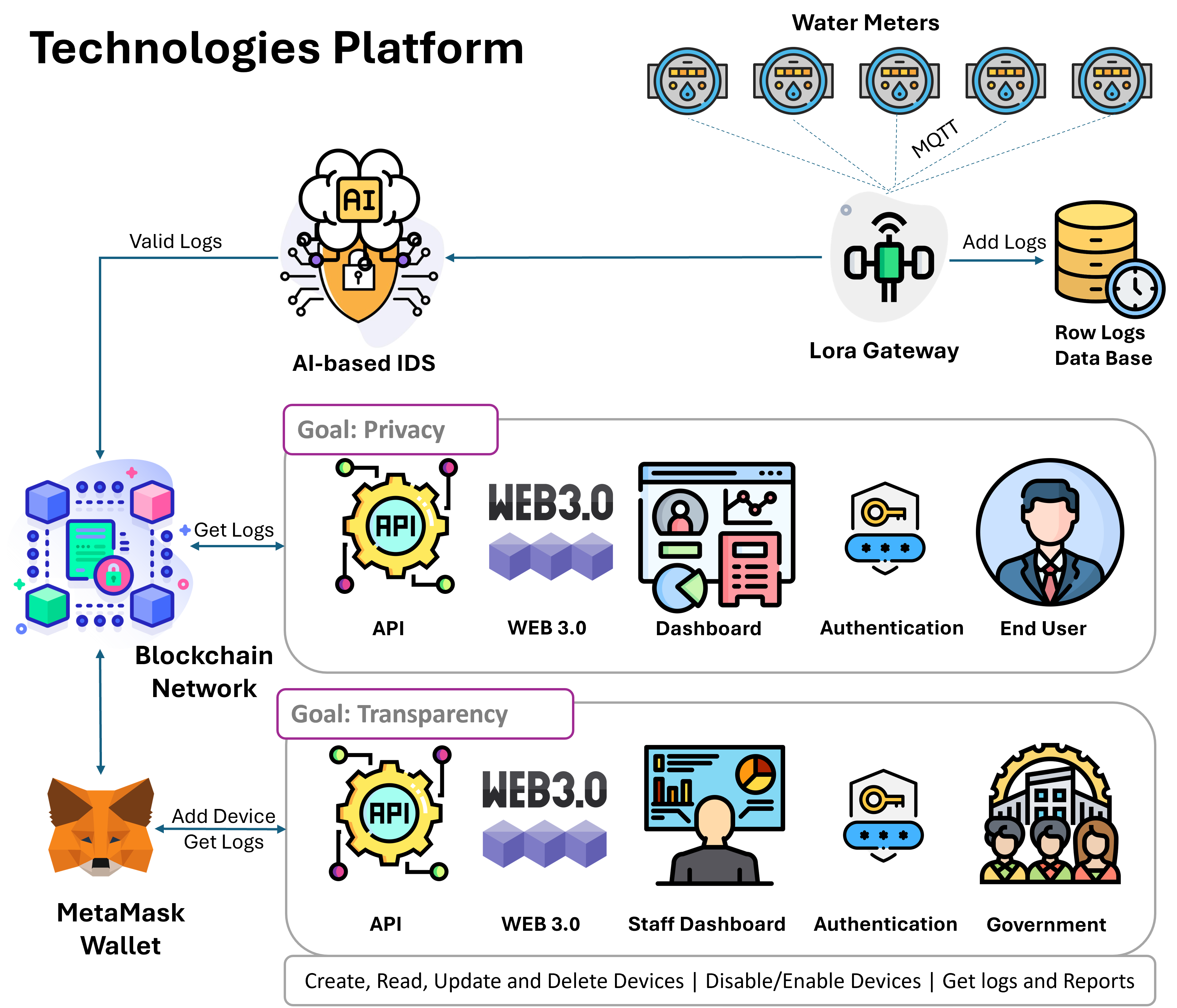}
    \caption{Technologies in the Platform on the BC side}
    \label{fig:blockchain-interface}
\end{figure}

In parallel, Figure~\ref{fig:blockchain-interface} presents the core technologies involved in the BC layer. This includes the use of a private Ethereum network with a PoA consensus model, which ensures fast finality and minimal energy consumption, particularly suitable for edge computing scenarios in rural environments. The integration of DTs with smart contracts enables not only secure data storage but also autonomous system behavior, reducing reliance on central servers or human intervention. By combining BC, DT, and AI-driven verification mechanisms, the platform offers a scalable, transparent, and resilient solution for water resource management in under-connected areas.

\section{Evaluation}\label{secIV}
This section presents a practical evaluation of the proposed framework under conditions typical of rural Spanish villages. We assess system-level performance—including throughput, latency, and scalability—as well as security and deployment cost. A private Ethereum blockchain with PoA consensus was deployed on a dedicated Hetzner server, and a LoRaWAN metering environment was used to emulate real-time consumption data. The evaluation includes analysis of the hardware/software setup, network topology, anomaly detection, tamper resistance, and cost-efficiency.

\subsection{Leakage Detection Results}
To verify the effectiveness of the leakage detection mechanism, we analyzed historical water meter data from 400 devices deployed across rural locations for last three years. The detection algorithm flagged meters with non-zero night consumption over consecutive days, suggesting probable leaks.

\begin{figure}[htbp]
    \centering
    \includegraphics[width=1\linewidth]{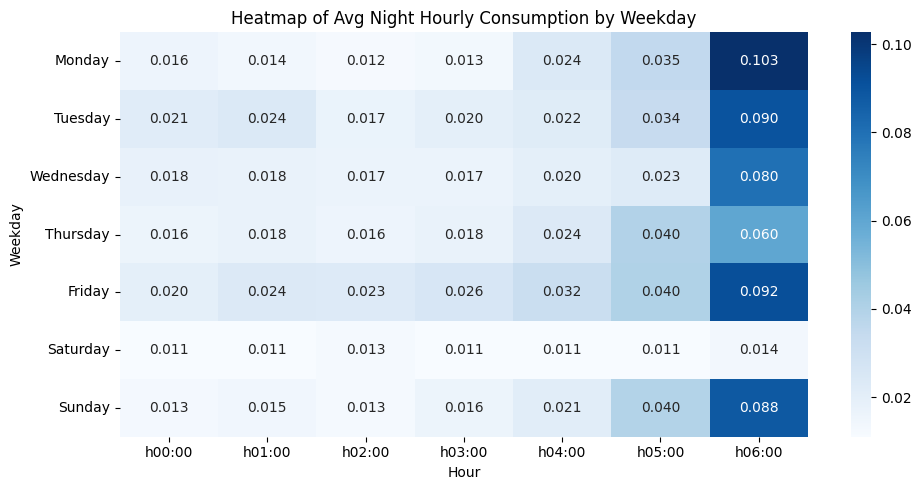}
    \caption{Night consumption Hitmap for a water meter with leakage}
    \label{fig:nightconsumption}
\end{figure}

\begin{figure}[htbp]
    \centering
    \includegraphics[width=1\linewidth]{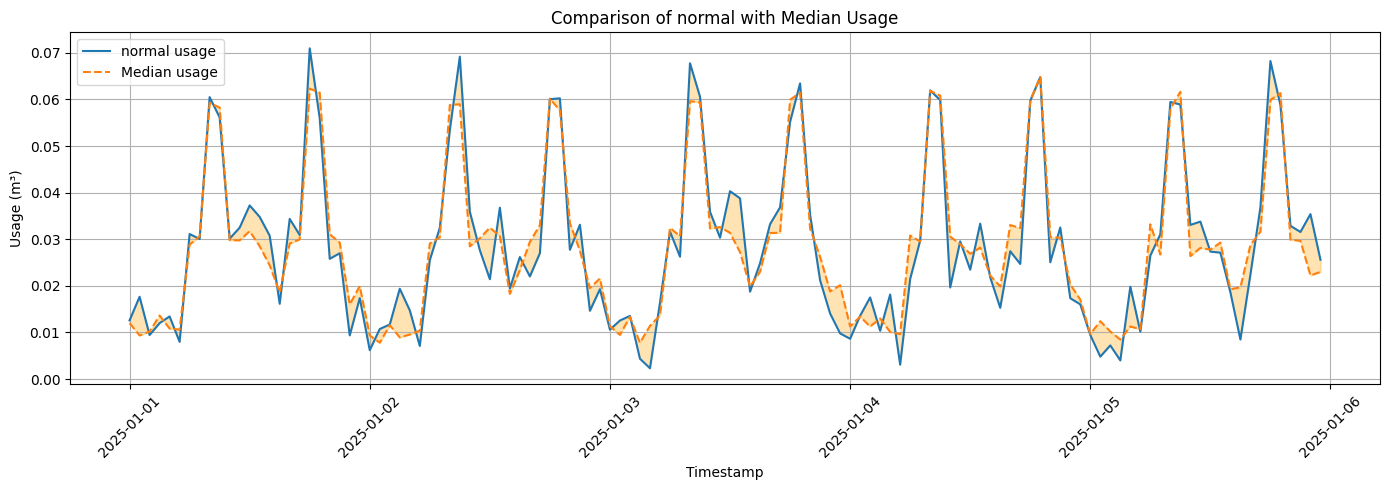}
    \caption{Comparison Normal and Median usage}
    \label{fig:Normal}
\end{figure}

\begin{figure}
    \centering
    \includegraphics[width=1\linewidth]{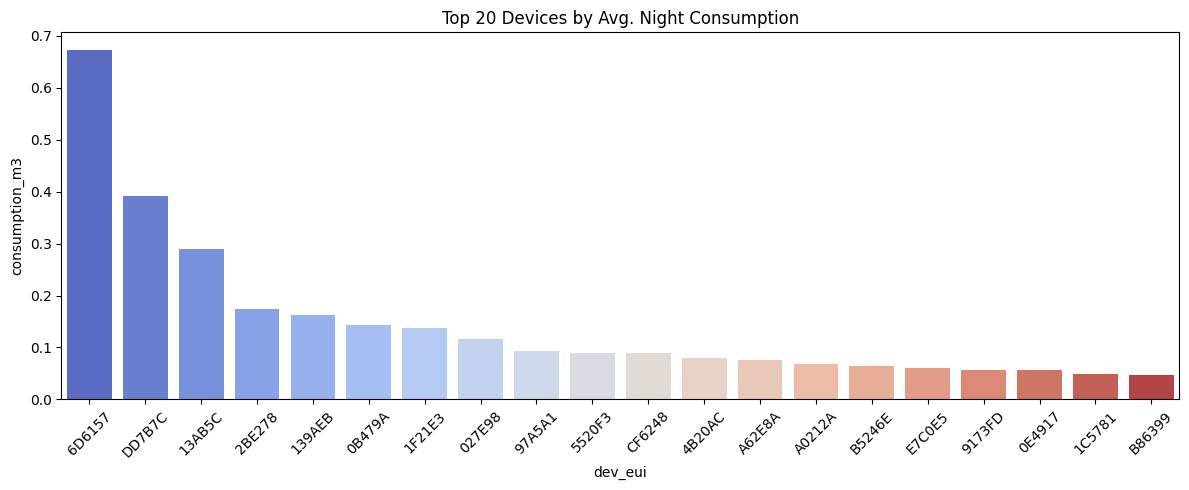}
    \caption{Leaked meters and their consumption during nights}
    \label{fig:leaked meter}
\end{figure}

Figure~\ref{fig:nightconsumption} shows a heatmap of night-time water consumption (00:00–06:00) for a leaking meter, where consistent activity was detected.  Figure~\ref{fig:Normal} compares normal consumption patterns to the median usage, highlighting that anomalies often deviate from expected seasonal or diurnal trends.  
Additionally, Figure~\ref{fig:leaked meter} aggregates the night usage across all flagged meters, reinforcing the accuracy of the detection logic.

\subsection{Anomaly Detection Result via IDS}

\begin{figure} [htbp]
\centering 
\includegraphics[width=1\linewidth]{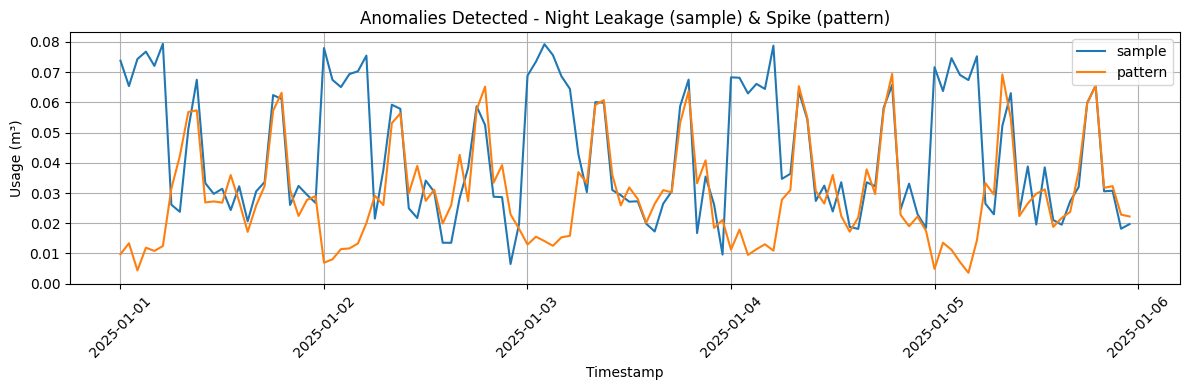} \caption{Anomalies Detected: Comparison of Night Leakage (sample) and Spike Patterns} 
\label{fig:Anomaly} 
\end{figure}

 The system was evaluated using real-world water consumption data injected with synthetic attacks. Figure~\ref{fig:Anomaly} shows a direct comparison between a normal consumption pattern and an anomalous one. The sample line represents typical usage behavior, while the pattern line reveals injected anomalies, such as abnormal spikes and repeated low-consumption values that mimic night-time leakage or spoofed records.

\begin{figure}[htbp]
\centering
\includegraphics[width=1\linewidth]{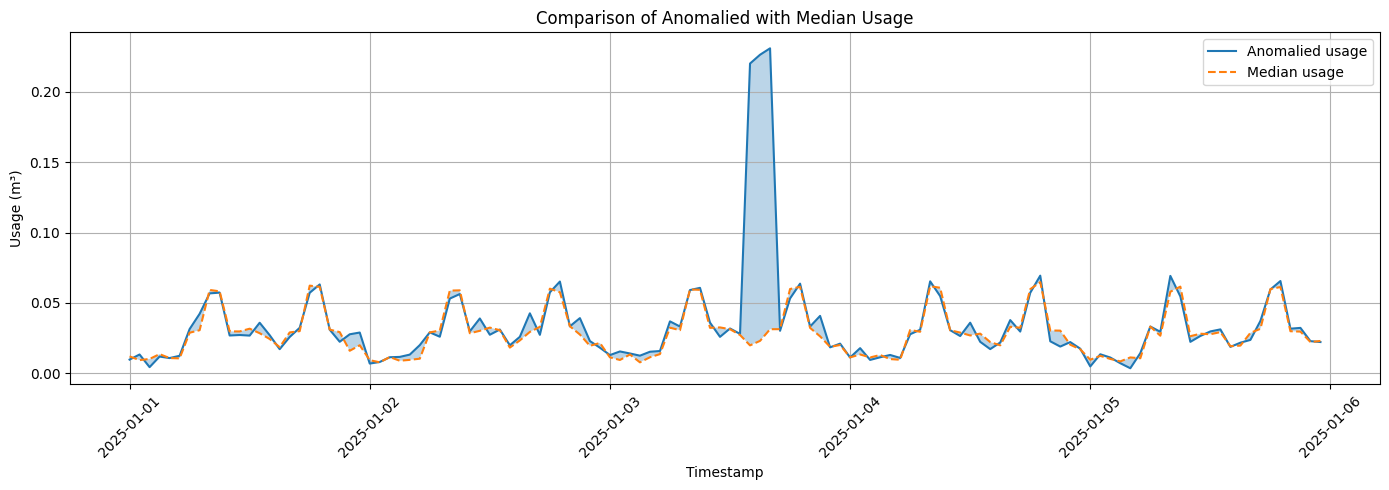} \caption{Detected Anomalies Compared with Median Usage Baseline} \label{fig:anomalyvsmedian}
\end{figure}

To further emphasize deviations, Figure~\ref{fig:anomalyvsmedian} plots the detected anomalous consumption values against the meter’s typical median usage. The sharp divergence observed during the attack period clearly demonstrates how the IDS identifies data points that deviate significantly from normal patterns while maintaining temporal consistency.

\begin{figure}[htbp]
\centering
\includegraphics[width=1\linewidth]{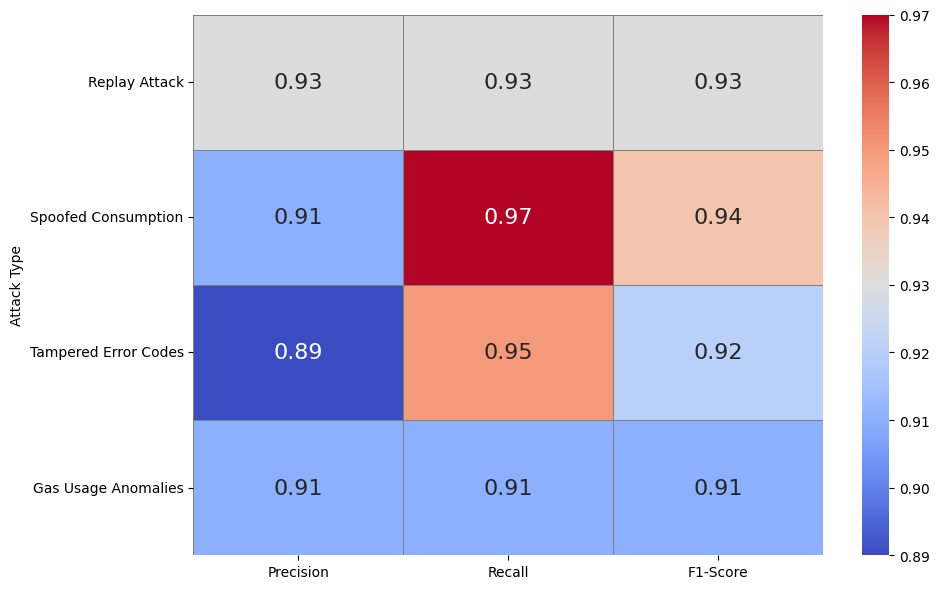}
\caption{Heatmap of Precision, Recall, and F1-Score per Attack Type}
\label{fig:Hitmapattack}
\end{figure}

Figure~\ref{fig:Hitmapattack} presents a heatmap comparing the IDS performance metrics—Precision, Recall, and F1-Score—for each type of attack. This visual summary is consistent with the quantitative results shown in Table~\ref{tab:ids_results}, highlighting the IDS’s effectiveness across diverse anomaly types.

\begin{table}[htbp]
\caption{Anomaly Detection Results of the Hybrid IDS}
\centering
\resizebox{8.8cm}{!}{
\begin{tabular}{lccccc}
\hline
\textbf{Attack Type} & \textbf{Injected} & \textbf{Detected} & \textbf{Precision} & \textbf{Recall} & \textbf{F1-Score} \\
\hline
Replay Attack         & 120 & 112 & 0.93 & 0.93 & 0.93 \\
Spoofed Consumption   & 100 & 97  & 0.91 & 0.97 & 0.94 \\
Tampered Error Codes  & 80  & 76  & 0.89 & 0.95 & 0.92 \\
Gas Usage Anomalies   & 70  & 64  & 0.91 & 0.91 & 0.91 \\
\hline
\textbf{Overall}      & 370 & 349 & \textbf{0.91} & \textbf{0.94} & \textbf{0.92} \\
\hline
\end{tabular}
}
\label{tab:ids_results}
\end{table}

\subsection{Experimental Setup BC}
\begin{itemize}
    \item \textbf{Hardware and Software Configuration:}
    
 The proposed framework was deployed on a Hetzner server running a private Ethereum network with a PoA consensus mechanism. Three validator nodes operated via Docker containers to simulate on-chain validation. Meter data from 400 emulated LoRaWAN devices was batched and submitted in 8-hour intervals. Prometheus and Grafana were used for monitoring. The setup reflects real-world rural conditions, tolerating intermittent connectivity while ensuring secure, scalable, and low-latency operation.

    \item \textbf{Network Topology:} 
    \begin{itemize}
        \item \textit{Validator Nodes:} Each validator node runs with a 1-second block interval and a block gas limit of 15 million—settings that enable higher throughput and faster transaction finalization than default Ethereum configurations. We deployed three validator nodes on the same Hetzner dedicated server, each operating in its own Docker container. The nodes communicate over a secure internal network and use static peer discovery to maintain connectivity and validate on-chain transactions.  

    \item \textit{LoRaWAN Gateway:} Meter data is transferred to the BC through a replicated LoRaWAN gateway that aggregates sensor data every 8 hours. It then batches these readings into transactions before submitting them on-chain, ensuring any intermittent connectivity does not result in data loss.
   \end{itemize}
    \item \textbf{Water Metering Scenario:}  
    In our test scenario, we emulate the behavior of 400 water meters installed across multiple rural areas. Each meter is configured to capture the following data at 8-hour intervals:
    \begin{itemize}
      \item Meter ID 
      \item Timestamp 
      \item Water Consumption (in cubic metres) 
      \item Error Code
    \end{itemize}
    Each meter transmits its data three times per day to account for potential downtime or connectivity issues. This setup reflects the actual operational conditions in remote Spanish villages, where consistent Internet access cannot always be guaranteed.
\end{itemize}

\subsection{Performance and Scalability}\label{sec:performance-scalability}
In this section, we present the results of our performance analysis and scalability tests for the proposed BC-based DT framework. We aim to demonstrate that the system can process meter readings efficiently, maintain low transaction latency, and scale to accommodate an increasing number of water meters.

\begin{itemize}

\item \textbf{Transaction Throughput and Latency:}

To evaluate the performance of the BC network, we focused on two core metrics:
\begin{itemize}
    \item \textit{ TPS's Throughput:} The number of successfully confirmed transactions per second.
    \item \textit{Transaction Latency (Seconds):} The time interval between the client submitting the transaction and its final confirmation on-chain.
\end{itemize}

We conducted a series of stress tests by varying the batch sizes (i.e., how many meter readings are grouped into a single on-chain transaction). This approach allowed us to evaluate the system’s behavior under different data aggregation strategies—especially relevant for rural deployments where intermittent connectivity may lead to buffered uploads.

The Mean Latency in Table~\ref{tab:throughput_latency} refers to the time from when a transaction is submitted to its first inclusion in a block. Finality in our PoA setup typically arrives 1–2 blocks after inclusion, corresponding to an additional 2–3 seconds under typical loads.

\begin{table}[htbp]
\caption{Transaction Throughput and Latency under Different Batching Conditions}
\centering
\resizebox{8.8cm}{!}{
\begin{tabular}{ccccc}
\hline
\textbf{Batch Size} & \textbf{Meters Tested} & \textbf{Throughput (TPS)} & \textbf{Mean Latency (s)} & \textbf{Max Latency (s)} \\
\hline
1 reading/tx   & 400 & 110 & 1.2 & 2.1 \\
5 readings/tx  & 400 & 96  & 1.5 & 2.4 \\
10 readings/tx & 400 & 89  & 1.7 & 2.8 \\
20 readings/tx & 400 & 81  & 2.1 & 3.5 \\
\hline
\end{tabular}
}
\label{tab:throughput_latency}
\end{table}

\textit{Observations:}
\begin{itemize}
    \item As the batch size increases, throughput decreases slightly, attributable to larger transaction payloads requiring more on-chain processing time.
    \item Latency grows proportionally with the batch size. However, even at 20 readings per transaction, the network sustains an average throughput of 81 TPS with a mean latency of around 2 seconds.
    \item These results indicate that our PoA network can efficiently handle data bursts from hundreds of meters, making it suitable for real-world deployments where large numbers of sensors may periodically transmit readings.
\end{itemize}

\item \textbf{Block Finalization:}

Using a PoA consensus mechanism provides faster block finalization times compared to traditional Proof of Work (PoW) networks. Our experiments show that blocks are typically finalized within \textbf{2--3 seconds} under the tested workloads. This quick finality has two primary benefits:
\begin{enumerate}
    \item \textit{Timely Data Recording:} Water consumption data are confirmed on-chain almost immediately, enabling near real-time monitoring within the DT environment.
    \item \textit{Predictive Maintenance:} Rapid confirmation aids anomaly detection algorithms in swiftly identifying irregularities (e.g., leaks or sensor malfunctions), reducing response times and potential water losses.
\end{enumerate}

Such short block finalization intervals are particularly valuable for rural water management, where operators rely on accurate, up-to-date information to schedule maintenance tasks, plan usage patterns, and optimize resources.

\item \textbf{Scalability with Increasing Meter Count:}

To assess how the system behaves under a growing number of sensors, we conducted additional tests by gradually scaling the number of simulated meters from 100 to 1,000 while holding the transaction batch size at five readings per transaction. Across these tests:
\begin{itemize}
    \item The network maintained a throughput greater than 85 TPS in all experiments, even as the meter count increased by an order of magnitude.
    \item System latency showed minimal growth, reinforcing the notion that the PoA-based framework scales well with increased demand.
    \item These results underscore the system’s potential to extend to larger water distribution networks without significant performance degradation, making it suitable for both small rural communities and larger municipal deployments.
\end{itemize}

\end{itemize}

\subsection{Security and Reliability}\label{sec:security-reliability}
Security and reliability are central pillars of any data management solution for critical infrastructure like water distribution. BC’s immutable ledger and PoA-based access controls work together to ensure that the system is tamper-resistant and fault-tolerant. Below, we detail the measures taken to protect against unauthorized modifications and to maintain network reliability.

\begin{itemize}

\item \textbf{Data Immutability and Tamper Resistance:}

One of the chief advantages of a BC-based solution is the \textit{immutability} of on-chain records. We conducted targeted tests to confirm that malicious attempts to alter data or inject bogus information would be rejected:

\begin{itemize}
    \item \textit{Direct Database Manipulation:} We tried modifying the raw on-chain data files stored in the local node’s directory. The PoA consensus nodes detected mismatched hashes, invalidating the altered data.
    \item \textit{Smart Contract Override:} We attempted to call special administrative functions like \textit{logWaterData} and \textit{disableMeter} without proper credentials. These calls were blocked by role-based access controls enforced at the smart contract level.
    \item \textit{Spurious Node Injection:} We introduced a rogue node with a manipulated ledger history, which the existing validator nodes refused to add to the network.
\end{itemize}

Table~\ref{tab:tamper_resistance_results} summarizes the outcome of these tests:

\begin{table}[htbp]
\caption{Tamper-Resistance Test Results}
\centering
\resizebox{9cm}{!}{
\begin{tabular}{ll}
\hline
\textbf{Attempted Attack} & \textbf{Result} \\
\hline
Direct on-disk data modification & Rejected (immutable ledger) \\
Unauthorized smart contract invocation & Rejected (access control) \\
Rogue validator introduction & Blocked (PoA authority management) \\
\hline
\end{tabular}
}
\label{tab:tamper_resistance_results}
\end{table}

 All unauthorized modifications were invalidated by the network’s consensus protocol, confirming that meter data remains tamper-proof once recorded on-chain. This reliability is crucial for building trust among municipalities, local water authorities, and end-users.

\item \textbf{Access Control and Authentication:}

Access control is enforced via smart contracts. Before a meter can submit data, it must be registered on-chain by an authorized administrator. We tested unauthorized submissions to assess whether the system would correctly reject them:
\begin{itemize}
    \item \textit{Fake Meter ID:} A transaction with an unregistered meter ID triggered an immediate rejection in the \textit{isMeterRegistered} function.
    \item \textit{Valid Meter ID, Incorrect Credential:} If the transaction was signed by a private key not recognized by the PoA nodes, the network discarded the transaction before it reached the contract logic.
\end{itemize}

These findings confirm that the framework effectively prevents unauthorized data entries and ensures only valid meter readings are integrated into the DT environment.

\item \textbf{Network Reliability in Rural Deployments:}

Rural Spanish villages often face intermittent Internet connectivity. Our solution, therefore, tolerates temporary offline periods without losing data integrity. We simulated a scenario with a 2-hour daily connectivity loss over a week:
\begin{itemize}
    \item The LoRaWAN gateway buffered the readings until the BC node was reachable.
    \item Upon reconnection, the pending transactions were submitted in batches.
    \item No BC reorganization occurred, as the PoA validators incorporated the newly arrived batches without conflict.
\end{itemize}

This experiment demonstrates the system’s resilience, confirming that brief outages do not compromise the integrity or completeness of stored data. Such robustness is essential for real-world deployments where continuous high-speed Internet is not always available.

\end{itemize}

\subsection{Cost Analysis}\label{sec:cost-analysis}
Although public BCs typically require gas fees for each transaction, our private PoA network could be configured to impose negligible or zero gas costs, significantly reducing financial overhead for municipalities. Our PoA nodes are hosted on a dedicated Hetzner server, with monthly costs ranging between €20 and €50, depending on the chosen plan. BC maintenance is also cost-effective, as PoA consensus eliminates CPU-intensive mining tasks, and validator nodes have minimal resource requirements beyond basic computation and storage. In terms of network traffic, LoRaWAN-to-BC communications incur only minor data charges, while on-chain transaction fees can be set to near zero, avoiding high per-transaction costs. The system's scalability ensures that adding additional meters does not significantly increase operational expenses since the same validator nodes can efficiently handle larger data volumes within tested limits. Furthermore, the architecture’s linear scalability ensures that even large-scale deployments remain affordable. Table~\ref{tab:cost_breakdown} provides an approximate breakdown of costs for a six-month pilot project, excluding expenses related to physical LoRaWAN gateways or sensors, which vary based on specific deployment needs.

\begin{table}[htbp]
\caption{Cost Estimation Breakdown in a 6-Month Pilot}
\begin{center}
\resizebox{9cm}{!}{ % Resize dynamically to exactly 8 cm width
\begin{tabular}{|c|c|c|}
\hline
\textbf{Component} & \textbf{Cost (EUR)} & \textbf{Notes} \\
\hline
Server Rental (6 months) & 120--300 & Depends on hosting plan \\
\hline
Maintenance & \textasciitilde 50 & Occasional reboots, software updates \\
\hline
Energy & Included & Covered by hosting service \\
\hline
LoRaWAN Gateways & Variable & Based on deployment size and hardware choice \\
\hline
On-chain Gas Fees & Near-zero & PoA network with custom gas price \\
\hline
\end{tabular}
}
\label{tab:cost_breakdown}
\end{center}
\end{table}

Overall, this PoA-based BC solution is cost-effective for municipalities of varying sizes, especially when compared to traditional centralized data management systems that may involve higher maintenance and licensing fees.

\section{Conclusion}\label{secV}
This paper introduced a Blockchain-based Digital Twin (BC-DT) framework that combines a private PoA Ethereum blockchain, LoRaWAN sensors, and a hybrid Intrusion Detection System using LSTM Autoencoder and Isolation Forest to enhance the security and reliability of rural water distribution systems. The proposed system enables real-time anomaly detection, secure data logging via smart contracts, and supports transparent, decentralized monitoring. Evaluation results demonstrated strong performance with over 80 TPS, low latency, tamper resistance, and cost-effective scalability across 1,000 smart meters. The architecture is resilient to intermittent connectivity and adaptable to rural infrastructure constraints. Future work will explore enhancements such as federated learning for decentralized model training, dynamic pricing via smart contracts, and energy-efficient scaling to urban and industrial settings.

\section*{Acknowledgment}
This initiative is carried out within the framework of the funds from the Recovery, Transformation, and Resilience Plan, financed by the European Union (Next Generation) – National Institute of Cybersecurity (INCIBE), as part of project C107/23:
"Artificial Intelligence Applied to Cybersecurity in Critical Water and Sanitation Infrastructures."

% \bibliographystyle{IEEEtran} % Use IEEE style
% \bibliography{bibliography}  % Link to your .bib file (without .bib extension)

\end{document}